\documentclass[aps,prc,twocolumn,superscriptaddress,groupedaddress]{revtex4}
\usepackage{graphicx}
\usepackage{dcolumn}
\usepackage{bm}
\usepackage{amssymb}
\usepackage{amsmath}
\usepackage{color}

\newcommand{\bmd}{$\beta^{-}$-decay}
\newcommand{\bdne}{$\beta^{-}$-delayed neutron emission}
\newcommand{\bdf}{$\beta^{-}$-delayed fission}
\newcommand{\bdfs}{$\beta df$}
\newcommand{\spfs}{$sf$}
\newcommand{\nifs}{$(n,f)$}

\begin{document}

\hspace{5.2in} \mbox{LA-UR-20-24018}

\title{Following nuclei through nucleosynthesis: a novel tracing
technique}

\author{T.M. Sprouse\footnote{Los Alamos Center for Space and Earth
Science Student Fellow}}
\affiliation{Department of Physics, University of Notre Dame, Notre
Dame, IN 46556, USA}
\affiliation{Theoretical Division, Los Alamos National Laboratory, Los
Alamos, NM 87545, USA}

\author{M.R. Mumpower}
\affiliation{Theoretical Division, Los Alamos National Laboratory, Los
Alamos, NM 87545, USA}
\affiliation{Center for Theoretical Astrophysics, Los Alamos National
Laboratory, Los Alamos, NM 87545, USA}

\author{R. Surman}
\affiliation{Department of Physics, University of Notre Dame, Notre
Dame, IN 46556, USA}

\date{\today}

\begin{abstract}
Astrophysical nucleosynthesis is a family of diverse processes by which
atomic nuclei undergo nuclear reactions and decays to form new nuclei.
The complex nature of nucleosynthesis, which can involve as many as tens
of thousands of interactions between thousands of nuclei, makes it
difficult to study any one of these interactions in isolation using
standard approaches. In this work, we present a new technique,
nucleosynthesis tracing, that we use to quantify the specific role of
individual nuclear reaction, decay, and fission processes in
relationship to nucleosynthesis as a whole. We apply this technique to
study fission and {\bmd} as they occur in the rapid neutron capture
($r$) process of nucleosynthesis.
\end{abstract}

\maketitle

\section{Introduction}\label{sec:Introduction}

The extreme conditions that can arise in astrophysical environments
enable nuclear transmutation processes to take place, by which atomic
nuclei interact with their environment or decay to form new nuclei.
Insofar as different astrophysical environments may foster certain
transmutation processes but not others, these environments may be
categorized by the different types of nucleosynthesis that occur in
each; one of the primary goals of nuclear astrophysics, then, is to
explain how these different nucleosynthesis sources produce all of the
chemical elements observed in the universe, beginning with the
primordial hydrogen and helium produced during the Big Bang \cite{b2fh}.

In the most complex cases, nucleosynthesis can involve many thousands of
nuclear species connected by upward of $\sim$100,000 nuclear
transmutation processes by which their abundances may evolve in time.
Because the rates at which the different processes occur may depend on
the temperature and density of the environment in which the nuclei are
situated, as well as the abundances of the different nuclei themselves,
nucleosynthesis is an extremely dynamical and nonlinear problem.
Nevertheless, through the use of nuclear reaction networks, it is
possible to effectively model nucleosynthesis numerically.

More difficult, however, is the problem of isolating and quantifying the
role of individual nuclear properties as they influence nucleosynthesis
as a whole. This can be especially important because nucleosynthesis is
inherently determined by the properties of the nuclei that participate
in it, making these properties the focus of a significant number of
experimental and theoretical campaigns in nuclear physics (see, e.g.,
\cite{Wallerstein1997,Bertulani2010,Rauscher2010,Horowitz2019,Kajino2019}
and references therein). By identifying the most crucial nuclear
properties for each nucleosynthesis process, these campaigns may be
more precisely focused in the near-future on those properties which will
most significantly constrain nucleosynthesis simulation predictions.

Past approaches to accomplish this goal have either (1) systematically
varied individual or collections of nuclear properties and examined the
relative changes to nucleosynthesis, for example in
\cite{Schatz1998,Iliadis2002,
Mumpower+2016,Surman+2016,Martin+2016,
Bliss2017,Denissenkov2018,Sprouse+DFT}
or (2) analyzed the overall rate at which different processes
(reactions, decays, or fission) occur during nucleosynthesis, such as in
\cite{Wanajo+2007,Mumpower+2018,Vassh+2019}. However, it has not been
possible using current techniques to precisely quantify which nuclei,
and in what amounts, are affected by individual nuclear properties.
Approach (1) inherently modifies the nucleosynthesis simulation itself,
insofar as decay and reaction rates themselves are modified. While
approach (2) does not affect nucleosynthesis simulations in the same
way, it provides only limited insight into which aspects of a simulated
abundance pattern are affected.

In this work, we introduce a new framework for nucleosynthesis modeling,
\textit{nucleosynthesis tracing}, that can be applied to supplement
these two approaches. Nucleosynthesis tracing enables the robust
quantification of which nuclei have participated in an arbitrary
collection of nuclear reactions, decays, and/or fission at some point
during nucleosynthesis. In Sec.~\ref{sec:theory}, we identify the
underlying assumptions of the nucleosynthesis tracing framework and
derive the differential equations that define the technique. In
Sec.~\ref{sec:PRISM}, we briefly summarize how we have implemented
nucleosynthesis tracing as PRISM\textsuperscript{tr}, a modified version
of the nuclear reaction network code Portable Routines for Integrated
nucleoSynthesis Modeling (PRISM). Finally, we demonstrate several
possible applications of nucleosynthesis tracing to
rapid neutron capture ($r$ process) nucleosynthesis in
Sec.~\ref{sec:apps}.

\section{Theory}\label{sec:theory}

Traditional nuclear reaction networks time-evolve the nuclear abundances
of a system actively undergoing nucleosynthesis. These calculations
require a number of different input parameters, including an initial
composition of nuclei and any relevant nuclear properties, such as
nuclear reaction and decay rates. Because these rates may depend on the
thermodynamics of the system, it is also necessary to specify and evolve
the temperature and density during nucleosynthesis. Reaction networks
may also incorporate other environmental properties, such as external
heating rates and (anti)neutrino fluxes, into their calculations.

The generalized problem of simulating nucleosynthesis may then be
phrased as follows. We assume that each nuclide in a system can be
uniquely identified by its proton number, $Z,$ and mass number, $A.$ To
each combination of these $(Z,A)$, we assign an integer, $i,$ that
indexes the species. The number density of species $i$ at any given time
is given as $n_i,$ and we define the relative abundance of this species
as $Y_i = n_i/\rho N_A$, where $\rho$ is the baryon density, and the
mass fractions $X_i=Y_iA_i$ sum to 1, $\sum_i X_i=1$.

Changes in abundances are enacted by a collection of nuclear
transmutation processes, $P.$ For each process $p$ in $P,$ we require
the associated quantities listed below.
\begin{enumerate}
\item A rate function $\Lambda_p,$ which is allowed to depend explicitly
on any environmental quantity available to the network, such as
temperature, density, or neutrino flux, as appropriate to the process.
Insofar as each of these quantities is available to the network as a
function of time, the function $\Lambda_p$ is implicitly a function of
time.
\item A set of nuclear indices, $\mathcal{R}_p,$ that correspond to the
nuclei consumed by the process.
\item A set of nuclear indices, $\mathcal{P}_p,$ that corresponds to the
nuclei produced by the process.
\item A function, $\alpha_p(i),$ that gives the number of species with
index $i\in\mathcal{R}_p$ consumed by the process.
\item A function, $\beta_p(i),$ that gives the average number of species
with index $i\in\mathcal{P}_p$ produced by the process.
\end{enumerate}

Once a specific collection of relevant processes, $P,$ is specified,
nucleosynthesis is reduced to solving an initial value problem (IVP) for
the abundances as a function of time, $Y_i (t).$ The abundances at time
$t_0$ are taken as the initial condition. The set of processes, $P ,$
completely defines the system of differential equations for the IVP. For
each species in the network, $i,$ the differential equation defining its
evolution in time is given as
\begin{equation}\label{eq:traditional_net}
\begin{split}
\frac{dY_i}{dt} = & -\sum\limits_{ \{ p \in P \mid i \in \mathcal{R}_p
\} } \left( \alpha_p (i) \Lambda_p(t) \prod\limits_{j \in \mathcal{R}_p}
Y_j(t)^{\alpha_p(j)} \right) \\
& +\sum\limits_{ \{ p \in P \mid i \in \mathcal{P}_p \} } \left( \beta_p
(i) \Lambda_p(t) \prod\limits_{j \in \mathcal{R}_p}
Y_j(t)^{\alpha_p(j)}\right),
\end{split}
\end{equation}
where the first summation is taken over processes in which nuclide $i$
is given in $\mathcal{R}$ and the second over processes in which nuclide
$i$ is given in $\mathcal{P}.$ Because our notation differs
significantly from more commonly adopted forms, we refer to the Appendix
for a description of the relationship between
Eq.~\ref{eq:traditional_net} and that used, e.g., in \cite{Hix+1999}.

The specific approaches that are taken to solve this IVP vary across
reaction networks, and the ideal numerical methods can be
application-dependent. For simulating $r$-process nucleosynthesis, a
common approach is to solve an implicit Euler equation using the
Newton-Raphson method
\cite{Arnett+1969,Woosley+1973,Hix+1999,Hix+2006,Lippuner+2017},
although alternative approaches have also demonstrated success
\cite{Timmes+1999,Guidry+2012,Guidry+2013a,
Guidry+2013b,Guidry+2013c,Brock+2015}.

\subsection{Nucleosynthesis Tracing}
In this work, we develop an extension to reaction networks as previously
defined. We refer to this extension as a \textit{tracing} reaction
network. This extension, which is laid out below, enables the robust
quantification of which nuclei in a system have a nucleosynthetic
history involving a particular process or processes, among other
possible applications.

We begin by constructing a parallel set of abundances to evolve, denoted
as $Y_{traced,i}$ and referred to as the \textit{traced abundances.}
Physically, the traced abundances identify the subset of the total
abundances which have assumed some property, a \textit{trace-in
condition}, during nucleosynthesis. Once a nucleus satisfies the
trace-in condition, we add it to the traced abundances. We may also wish
to remove nuclei from consideration in the traced abundances after they
assume some other property, a \textit{trace-out condition}. We otherwise
evolve the overall abundances according to Eq.~\ref{eq:traditional_net}
and the traced abundances according to a slightly modified form of the
same equation. It is also helpful to define the \textit{untraced}
abundances as $Y_{untraced,i}=Y_{i} - Y_{traced,i}.$

For the present work, we strictly consider trace-in and trace-out
conditions to be the participation in a collection of processes,
$P_{\text{in}}$ and $P_{\text{out}},$ respectively. Effectively, we
begin tracing a nucleus once it is produced by a process identified in
$P_{\text{in}},$ and we continue to follow it throughout all subsequent
nucleosynthesis in which it participates. However, if it is consumed by
one of the processes in $P_{\text{out}},$ we remove it from the traced
abundances, and we do not consider any further nucleosynthesis in which
it participates. We also define an additional set of processes,
$P_{\text{other}},$ which we define as the set of those processes in $P$
belonging to neither $P_{\text{in}}$ nor $P_{\text{out}}.$

\begin{figure*}[!ht]
\begin{center}
\includegraphics[width=.750\textwidth]{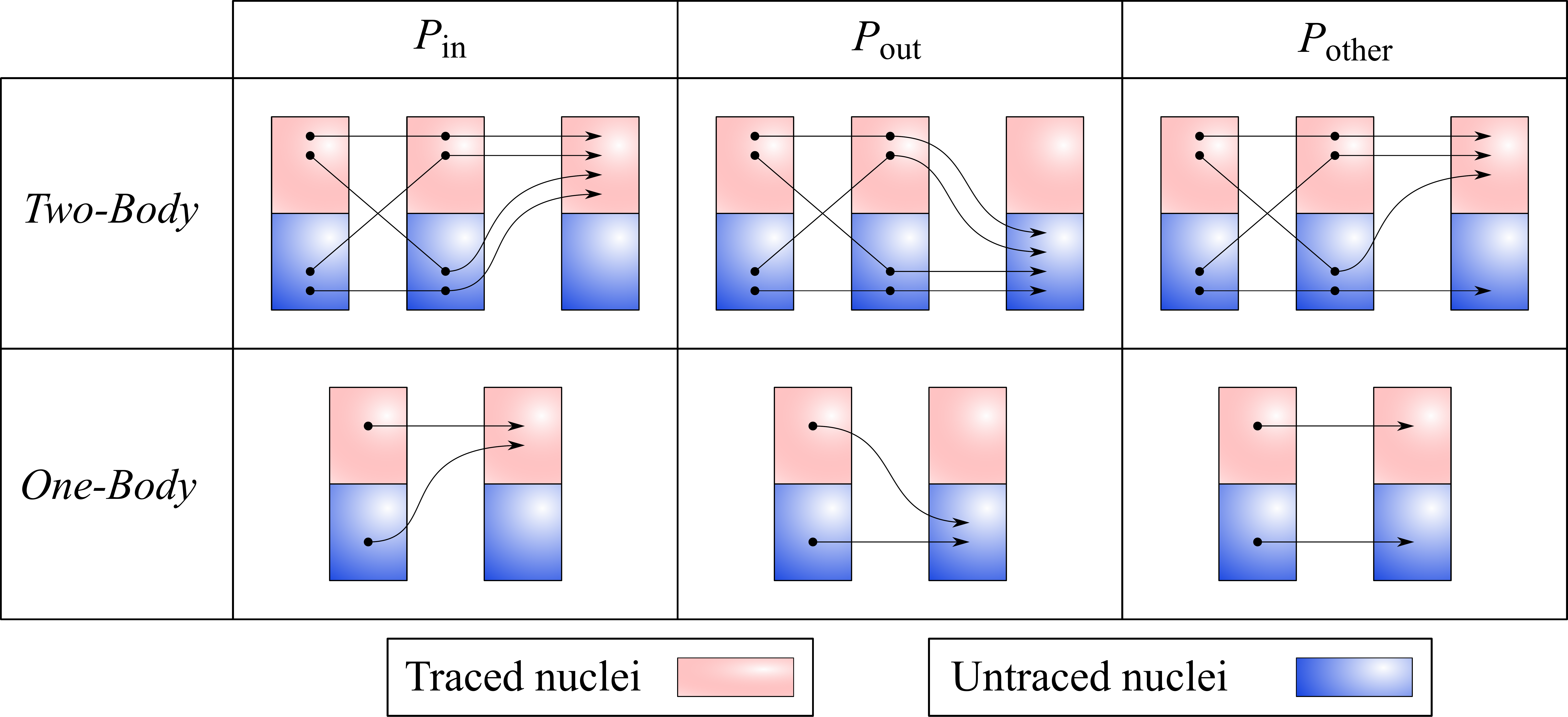}

\caption{\label{fig:tr_diag} A schematic diagram of nucleosynthesis
tracing. Populations of traced nuclei (top red boxes) and untraced
nuclei (bottom blue boxes) interact via one-body and two-body trace-in
processes ($P_{\text{in}}$), trace-out processes ($P_{\text{out}}$), and
other processes ($P_{\text{other}}$). The filled circles represent
nuclei consumed by a process, and the arrowheads represent nuclei
produced by a process. For the trace-in processes, products are mapped
exclusively to the traced population, while the trace-out processes map
products exclusively to the untraced population. All other processes are
allowed to map products to either the traced or untraced populations,
depending on the populations to which the reactants belong.}
\end{center}
\end{figure*}

We illustrate schematically how one- and two-body processes in each of
$P_{\text{in}},$ $P_{\text{out}},$ and $P_{\text{other}}$ affect the
evolution of the traced abundances in Fig.~\ref{fig:tr_diag}. Processes
in $P_{\text{in}}$ and $P_{\text{out}}$ are relatively straightforward,
as products are always mapped into the traced and untraced populations,
respectively. For one-body processes in $P_{\text{other}},$ the mapping
is straightforward as well, as products of traced nuclei are mapped into
the traced population, and products of untraced nuclei are mapped into
the untraced population.

For two-or-more-body processes, the situation is more complicated. If a
traced nucleus interacts with a traced nucleus, then the products
clearly should be mapped into the traced population. Likewise, if only
untraced nuclei undergo the process, then the products unambiguously
belong in the untraced population. It is also possible that some number
of untraced nuclei interact with some other traced nuclei, and we are
forced to choose what fraction of the products belong to the traced and
untraced populations. The simplest choice is to assert that if
\textit{any} nucleus consumed by the process is traced, then the
products are always mapped into the traced population, as can be seen in
the top-right panel of Fig.~\ref{fig:tr_diag}. This is the choice we
explore in the present work, although other meaningful choices are
possible.

The system of differential equations describing the evolution of the
traced abundances may now be defined under these assumptions. There are
four distinct varieties of terms that present themselves in these
differential equations, and we construct each of them in turn.

\textit{A traced nucleus, $i$, is produced by a process, $p,$ in
$P_\textit{in}.$} Because we wish to add \textit{all} of the nuclei
produced via this process into the traced network, the relevant term
should be identical to that of the total abundance $Y_i,$ namely
\begin{equation}\label{eq:prod_pin}
R_1 = \beta_p (i) \Lambda_p(t) \prod\limits_{j \in \mathcal{R}_p}
Y_j(t)^{\alpha_p(j)}.
\end{equation}   

\textit{A traced nucleus, $i$, is produced by a process, $p,$ in
$P_\textit{out}.$} Because none of the nuclei produced via this process
should be mapped into the traced network, this term is simply 0.

\textit{A traced nucleus, $i$, is produced by a process, $p,$ in
$P_\textit{other}.$} All nuclei produced by this process should be
mapped into the traced network \textit{unless} all of the nuclei
involved are untraced. The rate at which \textit{only} untraced nuclei
undergo the process is given by
\begin{equation}\label{eq:prod_pother_a}
R'_2 =\beta_p (i) \Lambda_p(t) \prod\limits_{j \in \mathcal{R}_p}
Y_{untraced,j}(t)^{\alpha_p(j)}.
\end{equation}
The rate at which traced nuclei are produced is the difference between
Eq.~\ref{eq:prod_pin} and Eq.~\ref{eq:prod_pother_a},
\begin{equation}\label{eq:prod_pother_b}
\begin{split}
R_2 = & \beta_p (i) \Lambda_p(t) \prod\limits_{j \in \mathcal{R}_p}
Y_j(t)^{\alpha_p(j)} \\ & - \beta_p (i) \Lambda_p(t) \prod\limits_{j \in
\mathcal{R}_p} Y_{untraced,j}(t)^{\alpha_p(j)}.
\end{split}
\end{equation}   
Expressed in terms of only traced abundances $Y_{traced,i}$ and overall
abundances $Y_i,$ this reduces to
\begin{equation}\label{eq:prod_pother}
\begin{split}
R_2 = & \beta_p (i) \Lambda_p(t) \prod\limits_{j \in \mathcal{R}_p}
Y_j(t)^{\alpha_p(j)} \\ & - \beta_p (i) \Lambda_p(t) \prod\limits_{j \in
\mathcal{R}_p} \left[ Y_{j}(t)-Y_{traced,j}(t)\right]^{\alpha_p(j)}.
\end{split}
\end{equation}   
 
\textit{A traced nucleus, $i$, is consumed by a process, $p,$ in
$P_\text{in},$ $P_{\text{out}}$, or $P_{\text{other}}.$} A traced
nucleus may interact with any other nucleus, traced or otherwise, to
undergo a particular process. As such, the rate at which a traced
nucleus is consumed by the process will be in proportion to the overall
rate in the ratio $Y_{traced,i}$/$Y_{i}.$ The term is given by
\begin{equation*}
R_3 = - \frac{Y_{traced,i}}{Y_{i}} \left( \alpha_p (i) \Lambda_p(t)
\prod\limits_{j \in \mathcal{R}_p} Y_j(t)^{\alpha_p(j)} \right).
\end{equation*}
Because an abundance $Y_i$ may be $0,$ we rearrange this slightly as
\begin{equation}\label{eq:reac_pall}
\begin{split}
R_3 = - \alpha_p (i) \Lambda_p(t) Y_{traced,i} Y_{i}^{\alpha_p(i)-1}
\prod\limits_{j \neq i \in \mathcal{R}_p} Y_j(t)^{\alpha_p(j)}.
\end{split}
\end{equation}   

The system of equations that govern the traced reaction network is a
linear sum of all appropriate terms of the forms Eq.~\ref{eq:prod_pin},
Eq.~\ref{eq:prod_pother}, and Eq.~\ref{eq:reac_pall}, together with the
system of equations defined in Eq.~\ref{eq:traditional_net}. It is given
by
\begin{eqnarray}
\small
\begin{split} \label{eq:dydt}
\frac{dY_i}{dt} = - \sum\limits_{ \{ p \in P \mid i \in \mathcal{R}_p \}
} \left( \alpha_p (i) \Lambda_p(t) \prod\limits_{j \in \mathcal{R}_p}
Y_j(t)^{\alpha_p(j)} \right) \\
+\sum\limits_{ \{ p \in P \mid i \in \mathcal{P}_p \} } \left( \beta_p
(i) \Lambda_p(t) \prod\limits_{j \in \mathcal{R}_p}
Y_j(t)^{\alpha_p(j)}\right) \\
\\
\end{split} \\
\small
\begin{split} \label{eq:traced_dydt}
\frac{dY_{traced,i}}{dt}  = 
- \sum\limits_{ \{ p \in P \mid i \in \mathcal{R}_p \} } \Big[ \alpha_p
(i) \Lambda_p(t) Y_{traced,i} Y_{i}^{\alpha_p(i)-1}\\
\times\prod\limits_{j \neq i \in \mathcal{R}_p}
Y_j(t)^{\alpha_p(j)}\Big] \\
+\sum\limits_{ \{ p \in P_{\text{in}} \mid i \in \mathcal{P}_p \} }
\Big[ \beta_p (i) \Lambda_p(t) \prod\limits_{j \in \mathcal{R}_p}
Y_j(t)^{\alpha_p(j)}\Big]\\
+\sum\limits_{ \{ p \in P_{\text{other}} \mid i \in \mathcal{P}_p \} }
\Big[ \beta_p (i) \Lambda_p(t) \prod\limits_{j \in \mathcal{R}_p}
Y_j(t)^{\alpha_p(j)}\\
- \beta_p (i) \Lambda_p(t) \prod\limits_{j \in \mathcal{R}_p} \left[
Y_{j}(t)-Y_{traced,j}(t)\right]^{\alpha_p(j)}\Big]
\end{split}
\end{eqnarray}

This extended system of equations can then be solved using the same
numerical techniques as for traditional network equations
(Eq.~\ref{eq:traditional_net}). Both the total and traced abundances
evolve mostly separate from each other, with connections between the two
mediated by the $P_{\text{in}}$ and $P_{\text{out}}$ processes. Note
that because the equations for $\frac{dY_i}{dt}$ are the same as in
Eq.~\ref{eq:traditional_net}, they are not affected in any way by the
extended network, and simulating the total abundances will not be
affected by using the extended network equations. However, the total
abundances do arise in the equations for the $\frac{dY_{traced,i}}{dt},$
and through this dependence the dynamic interactions of the traced
abundances with the total abundances are effectively captured.

\section{Nucleosynthesis and PRISM}\label{sec:PRISM}
For the demonstrated applications of the nucleosynthesis tracing
framework presented in Sec.~\ref{sec:apps}, we use an updated version of
the reaction network code PRISM
\cite{Mumpower+2017,Zhu+Cf254,Mumpower+2018} denoted as
PRISM\textsuperscript{tr}.
The extended network equations defining the evolution of the traced
abundances, summarized as Eq.~\ref{eq:traced_dydt}, are structurally
very similar to those of the total abundances, Eq.~\ref{eq:dydt}.
Numerically, we solve both the total abundances and the traced
abundances over a series of discrete timesteps by solving an implicit
Euler equation using the Newton-Raphson method. In addition to solving
for the time derivatives $\frac{dY_i}{dt}$ and
$\frac{dY_{traced,i}}{dt},$ this approach requires evaluating the
partial derivatives
\begin{align*}
 \frac{\partial}{\partial Y_j} & \left( \frac{dY_i}{dt} \right), \\
\frac{\partial}{\partial Y_j} & \left( \frac{dY_{traced,i}}{dt} \right),
\\
\frac{\partial}{\partial Y_{traced,j}} & \left( \frac{dY_i}{dt} \right),
\text{ and}\\
\frac{\partial}{\partial Y_{traced,j}} & \left( \frac{dY_{traced,i}}{dt}
\right).
\end{align*}
Insofar as Eqs.~\ref{eq:dydt}~and~\ref{eq:traced_dydt} are polynomials
of the $Y_i$ and $Y_{traced,i},$ these partial derivatives are
straightforward to evaluate, and we do not give their explicit form
here.

For the calculations performed in this work, we use
PRISM\textsuperscript{tr} to perform a number of $r$-process
nucleosynthesis tracing simulations. In all cases, we implement a
combination of experimental data and theory calculations for
charged-particle reaction rates; {\bmd} rates; delayed neutron emission
probabilities; neutron-capture rates; one-neutron photodissociation
rates; neutron-induced, $\beta^-$-delayed, and spontaneous fission
rates; and fission yields. Charged-particle reaction rate data is taken
from the JINA REACLIB database \cite{Cyburt2010}. The {\bmd} rates,
{\bdf} rates, and delayed neutron emission probabilities are evaluated
using the Los Alamos National Laboratory (LANL) QRPA+HF framework of
\cite{MumpQRPA+HF,MollerQRPA} using AME2016 \cite{wang2017} and FRDM2012
\cite{frdm2012} nuclear masses. Neutron-capture rates and
neutron-induced fission rates are calculated using the LANL statistical
Hauser-Feshbach code CoH \cite{Kawano+2016}, also assuming AME2016 and
FRDM2012 nuclear masses. One-neutron photodissociation rates are
evaluated by detailed balance, with the requisite one-neutron separation
energies taken from the AME2016 and FRDM2012 nuclear masses. Fission
yields are taken from the calculations of \cite{Mumpower2020}. We
implement all decay half-lives and branching ratios of the Nubase 2016
evaluation \cite{nubase2016}, which are taken to replace the
aforementioned theory calculations when possible.

Finally, we note that many of these processes invariably produce one or
more free neutrons. We do not, for the present work, intend to trace the
nucleosynthesis in which these neutrons participate. In order to prevent
such neutrons from populating the traced abundances, we fix
$\frac{dY_{traced,neutron}}{dt}=0,$ instead of evaluating it according
to Eq.~\ref{eq:traced_dydt}. Future work may investigate, e.g., the
relative effect of these neutrons on $r$-process nucleosynthesis, in
which case it would be necessary to evaluate
$\frac{dY_{traced,neutron}}{dt}$ via Eq.~\ref{eq:traced_dydt}.

\section{Applying the Nucleosynthesis Tracing Framework to the $r$
process}\label{sec:apps}
The rapid neutron capture process ($r$ process) of nucleosynthesis is
the astrophysical mechanism by which the heaviest elements observed to
exist in the universe are formed. The $r$ process proceeds by an
alternating sequence of neutron capture and {\bmd} towards progressively
heavier nuclei and is made possible by extremely hot, dense, and
neutron-rich environments \cite{b2fh}; identifying the astrophysical
events that provide such extreme conditions remains one of the greatest
open problems in nuclear astrophysics.
\cite{council2003,physicsTCAON2013}. Major progress towards this goal
occurred with the first gravitational wave observation of a neutron star
- neutron star merger, GW170817/GRB170817a/SSS17a \cite{GW170817+LIGO}.
Analysis of the electromagnetic counterpart of this event suggests a
significant lanthanide component in the ejecta of this event, pointing
to neutron star mergers as one possible site of the $r$ process
\cite{GW170817+EM}. However, owing to significant challenges in
observational astronomy, astrophysics, and nuclear physics, it is not
yet possible to clearly identify neutron star mergers as the dominant
source of $r$-process nuclei in the universe (see, e.g.,
\cite{Horowitz2019,Kajino2019} and references therein).

The $r$ process poses several barriers to analysis that make it a
particularly interesting focus for the first application of our tracing
framework. Multiple nuclear processes, including neutron capture,
neutron photodissociation, and {\bmd}, are all in competition as
thousands of different nuclear species are populated throughout
nucleosynthesis; when some of the heaviest and most neutron-rich nuclei
are formed, fission begins to compete as well, populating lighter nuclei
according to complex fission fragment distributions (yields) that
potentially span hundreds of different nuclei. Because of the large
number of nuclear species involved and the numerous transmutation
processes connecting them, it can be especially difficult to quantify
how individual transmutation processes interact with the many others to
determine the progression of nucleosynthesis.

In this section, we demonstrate several ways that our nucleosynthesis
tracing framework may be applied to address the challenges associated
with understanding the role of nuclear properties in governing
$r$-process nucleosynthesis. Sec.~\ref{sec:global} highlights the role
of different fission channels \textemdash considered as a whole, as well
as for individual nuclei \textemdash in a variety of neutron star merger
$r$-process environments. In Sec.~\ref{sec:betas} we choose a set of
neutron star merger wind conditions where fission plays a minimal role
and perform tracing calculations for the {\bmd} of elements $40\leq
Z\leq 80.$

\subsection{Distribution of fission products in final r-process
abundances}\label{sec:global}
In extremely neutron-rich environments, sufficiently heavy nuclei may be
formed during the $r$ process such that these nuclei begin to fission,
with varying degrees of significance for the nuclear abundances produced
from nucleosynthesis. In the most extreme cases, fission recycling may
occur, in which the bulk of nuclei undergo fission; these nuclei are
returned to lighter nuclei and undergo additional neutron captures and
{\bmd s} characteristic of the $r$ process. In such conditions, the
specific nuclear abundances are expected to depend on the fission
properties of many exotic neutron-rich nuclei
\cite{Beun2008,Korobkin2012,Goriely2015,Eichler2015,
Mumpower+2018,Vassh+2019,Vassh2019a}.
However, these fission properties effectively rely entirely on
theory-based predictions, and a great deal of progress has been made to
evaluate them, including from systematic, macroscopic-microscopic, and
purely microscopic theoretical approaches (see, e.g., \cite{Schmidt2018}
and references therein for a recent review; also
\cite{Giuliani2018,Lemaitre2018,Mumpower+2018,Lemaitre2019,
Vassh+2019,Mumpower2020,Witold2020}).
In order to help inform existing and future efforts in the study of
nuclear fission, we use our tracing framework to examine the various
ways that different fission processes, namely spontaneous fission
({\spfs}), {\bdf} ({\bdfs}), and neutron-induced fission ({\nifs}), can
influence $r$-process nucleosynthesis in the ejecta of a neutron star
merger.

We begin by considering a parameterized trajectory for the ejecta of a
neutron star merger accretion disk wind, with specific entropy
{$s/k_B=40$}, timescale {$\tau = 20$~ms}, and electron fraction {$Y_e =
0.20$}, as used in \cite{Zhu+Cf254}. Under these conditions, fission
plays a subdominant role that is insufficient to be characterized by
fission recycling. We perform three separate tracing calculations in
which the products of all nuclei which fission via a particular channel,
$(n,f)$, $\beta df$, or $sf$, are followed throughout the remaining
nucleosynthesis.

\begin{figure*}[!ht]
\begin{center}
\includegraphics{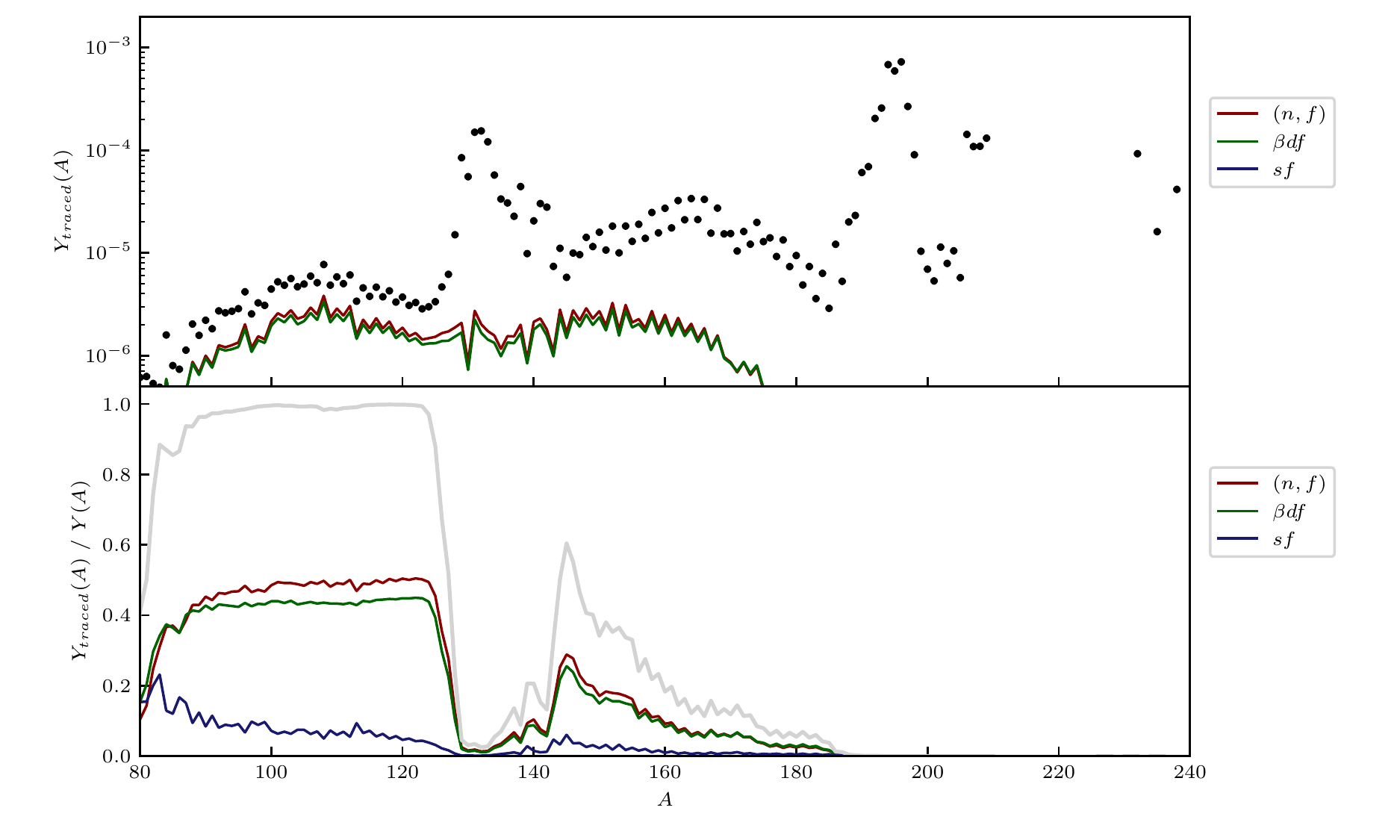}

\caption{\label{fig:global-CF20} Relative contributions to final
isotopic abundances by terminating fission channel (spontaneous fission,
$sf$; neutron-induced fission, $(n,f)$; and $\beta^-$-delayed fission,
$\beta df$) for the neutron star merger wind conditions described in the
text. The top panel compares the traced abundances (solid lines) to the
total abundances (dots). The bottom panel shows the ratio of each traced
abundance to the total abundance; the gray line indicates their sum. For
$A < 125$, all abundances are populated almost exclusively by fission
processes, while the relative contributions to $A > 125$ are
comparatively weaker, peaking at $\sim 50\%$. }
\end{center}
\end{figure*}

The results of each of these calculations are shown in
Fig.~\ref{fig:global-CF20}. The contributions are dominated by the
{\nifs} and {\bdfs} channels, with the final distribution of fission
products lying in the $80<A<180$ region. Among the fission contributions
to the overall pattern, roughly equal contributions arise from the
{\nifs} and {\bdfs} channels. In particular, we note that for these
conditions, very few nuclei remain to the left of the second $r$-process
peak ($A\sim130$) prior to the onset of fission, with contributions to
this region being dominated by material that is directly deposited there
as fission products. Because this fission occurs relatively late during
nucleosynthesis, the material does not significantly move forward into
the second or third ($A\sim195$) peaks via subsequent neutron capture,
as detailed in \cite{Mumpower+2018}.

We repeat this analysis for dynamical ejecta conditions from the
simulations of \cite{RW+Traj}. Fission is more significant in this case,
with around $60\%-80\%$ of heavy nuclei across the entire pattern having
participated in fission. Because most (but not all) of the heavy nuclei
have been processed through fission, we define this nucleosynthesis to
have proceeded via \textit{incomplete fission recycling}.

Significant amounts of {\nifs} products undergo follow-up
neutron-capture nucleosynthesis, forming up to $60\%$ of abundances well
beyond the extent of the fission yields, including the third $r$-process
peak and long-lived actinide isotopes. These contributions arise from
earlier stages of nucleosynthesis, when free neutrons are still
relatively abundant and extremely neutron-rich nuclei along the
$r$-process path undergo {\nifs}.

Nuclei to the left of the second peak are populated via a mechanism
similar to that of the wind conditions from Fig.~\ref{fig:global-CF20}.
Towards the end of nucleosynthesis, the free neutron abundance is
sufficiently low such that further neutron capture does not
significantly occur following fission. Because of this, significant
contributions in the $80<A<125$ region arise from the late-time {\nifs}
and {\bdfs} of less neutron-rich nuclei.

\begin{figure*}[!ht]
\begin{center}
\includegraphics{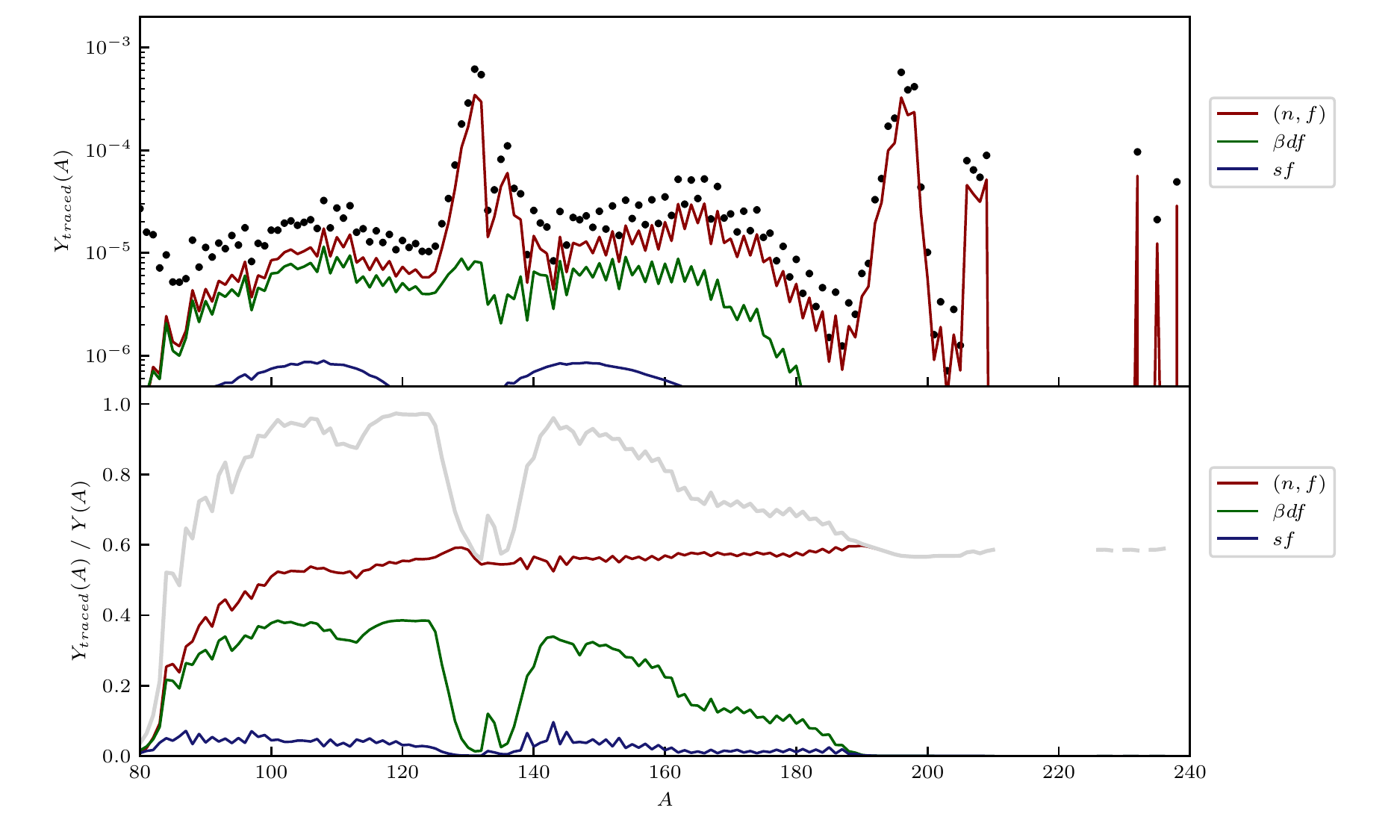}

\caption{\label{fig:global-RW17} Relative contributions to final
isotopic abundances by terminating fission channel (spontaneous fission,
$sf$; neutron-induced fission, $(n,f)$; and $\beta^-$-delayed fission,
$\beta df$) for dynamical ejecta conditions from a neutron star merger
simulation \cite{RW+Traj}, as in Fig.~\ref{fig:global-CF20}. For these
conditions, nucleosynthesis proceeds via incomplete fission recycling,
with neutron-induced fission accounting for $\sim 60\%$ of abundances
across the entire range of the pattern, and {\bdf} accounting for an
additional $10\%$ to $30\%$ for $A<\sim180$. The remaining abundances,
about $10\%$ to $40\%$ depending on $A,$ have no history of fission. }
\end{center}
\end{figure*}

\begin{figure*}[!ht]
\begin{center}
\includegraphics{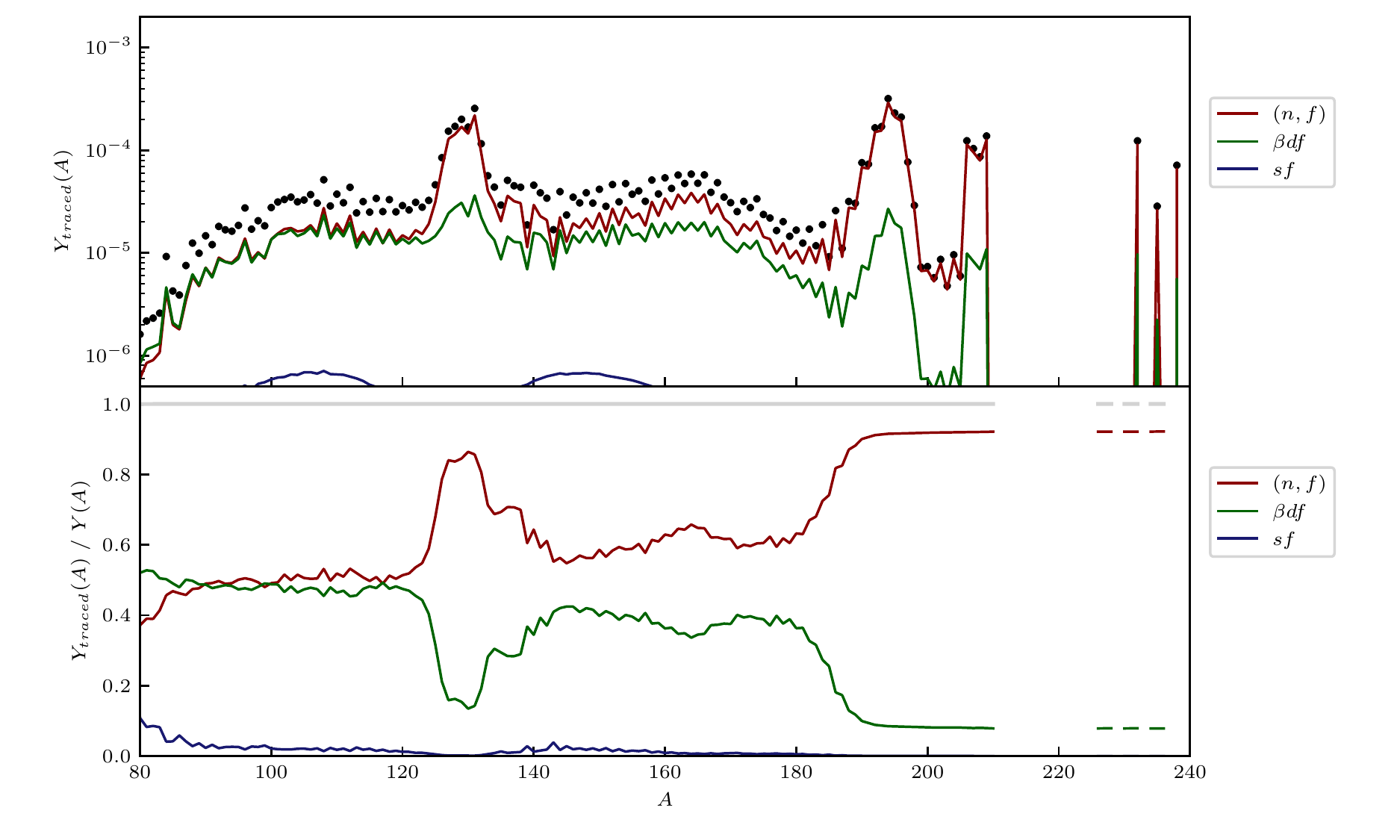}

\caption{\label{fig:global-RW1} Relative contributions to final isotopic
abundances by terminating fission channel (spontaneous fission, $sf$;
neutron-induced fission, $(n,f)$; and $\beta^-$-delayed fission, $\beta
df$) for the cold dynamical ejecta conditions from a neutron star merger
simulation \cite{RW+Traj}, as in Fig.~\ref{fig:global-CF20} and
Fig.~\ref{fig:global-RW17}. Under these conditions, nucleosynthesis
proceeds via complete fission recycling, with $100\%$ of the pattern
having a traced history involving $(n,f)$, $\beta df$, or $sf$. }
\end{center}
\end{figure*}

Finally, we consider cold tidal-tail conditions from \cite{RW+Traj}.
Here, all heavy nuclei have been processed one or more times through
fission, possibly via multiple fission channels. For our analysis, we
consider only contributions that arise from the terminating, or last,
fission event. In order to achieve this, we trace-in all fission events
for the particular channel under consideration and trace-out all other
fission events. If a particular abundance in the final pattern has a
history involving two different fission channels, only the contribution
from the last fission event is considered. As can be seen in
Fig.~\ref{fig:global-RW1}, the effect of this is that the sum of fission
traces across the {\nifs}, {\bdfs}, and {\spfs} channels add neatly to
unity, even though the average nucleus has more than one fission event
in its history. Insofar as all heavy nuclei have been processed at least
once through fission, we define this nucleosynthesis to have proceeded
via \textit{complete fission recycling}.

In contrast with the calculations shown in Fig.~\ref{fig:global-RW17},
{\bdfs} can become active earlier in nucleosynthesis, in competition
with {\nifs} for nuclei along the $r$-process path. As evidence of this,
note that the {\bdfs} products are able to undergo further neutron
capture reactions, eventually populating $\sim 10\%$ of the third peak
and long-lived actinide abundances, in addition to some movement of the
products from the $A<125$ region into the second peak. As with the
previous conditions, nuclei to the left of the second peak are dominated
by late-time fission products, with roughly equal contributions from the
{\nifs} and {\bdfs} channels.

While valuable insight can be derived from tracing entire fission
channels across all nuclei, it is also possible to apply our tracing
technique with much finer resolution by tracing the fission of
individual nuclei. While integrated fission flows have helped inform
which nuclei most actively undergo fission during the $r$ process (see
e.g. \cite{Mumpower+2018,Vassh+2019}), such approaches provide limited
information relating to the distribution of the fission products
throughout the abundance pattern at the conclusion of nucleosynthesis.
These effects become particularly important in conditions for which
nucleosynthesis proceeds via complete or incomplete fission recycling,
where there is a combination of early-time fission (whose products are
significantly reprocessed via neutron capture), late-time fission (whose
products are mostly restricted to {\bmd s} toward stable nuclei), and
intermediate cases.

\begin{figure*}[!ht]
\begin{center}
\includegraphics[width=0.85\textwidth]{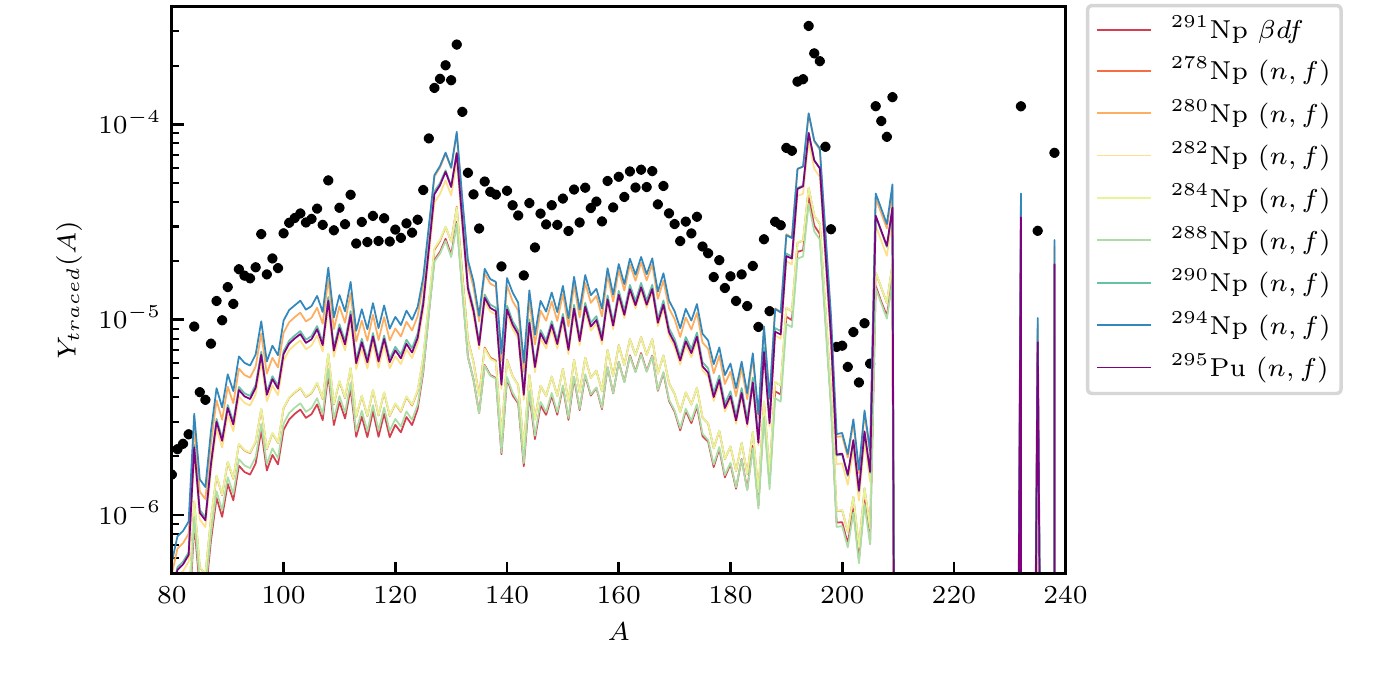}

\caption{\label{fig:rel-all} Relative contributions to final isotopic
abundances, ($Y(A),$ solid dots) for the $\beta^-$-delayed ($\beta df$)
and neutron-induced ($(n,f)$) fission products of individual neptunium
and plutonium isotopes (solid lines) in the cold dynamical ejecta
conditions of \cite{RW+Traj}, as in Fig.~\ref{fig:global-RW1}. The
plotted isotopes are those whose fission yields contribute at least
$10\%$ to the total abundance pattern for at least one mass number $A.$}
\end{center}
\end{figure*}

We perform tracing calculations for the {\nifs} and {\bdfs} of each
nuclide found to fission in the cold tidal-tail ejecta considered in
Fig.~\ref{fig:global-RW1} and appurtenant discussion. While $\sim 300$
nuclides fission via either channel during nucleosynthesis, we find that
the overall contribution of their products to the final calculated
abundances is quite small for the majority of these, on the order of
$1\%$ or less. By restricting to fission processes with traced
abundances constituting a minimum of $10\%$ of the final pattern for at
least one value of $A,$ we find $9$ fission processes to be significant
in these conditions, distributed across very neutron-rich neptunium
(Z=93) and plutonium (Z=94) isotopes. Collectively, these drive the
effects of early-time fission events presented in the discussion of
Fig.~\ref{fig:global-RW1}.

\begin{figure}[!ht]
\begin{center}
\includegraphics{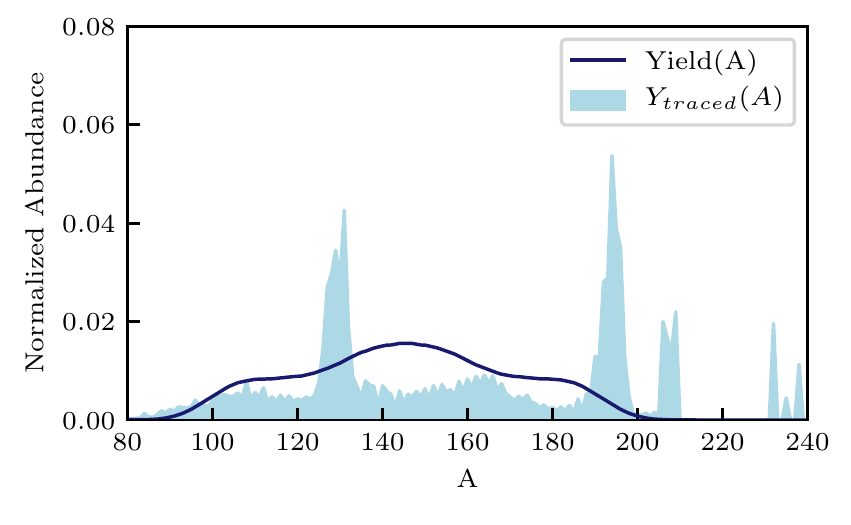}

\caption{\label{fig:yield-N-Np290} Comparison of fission yield (solid
line) to final traced abundances (shaded region) for the neutron-induced
fission of neptunium-290 in the cold dynamical ejecta conditions of
\cite{RW+Traj}, as in Fig.~\ref{fig:global-RW1}. Both are normalized
according to Eq.~\ref{eq:norm}. Nuclei produced by this fission process
participate in significant further neutron-capture nucleosynthesis, with
the actual fission yield leaving a minimal imprint on the final
abundance pattern.}
\end{center}
\end{figure}

In Fig.~\ref{fig:rel-all}, we plot the traced abundances for each of
these 9 fission processes. In each case, the traced abundances follow
the same shape as the total pattern, suggesting that the fission
products from these nuclei do not imprint on the final abundance
pattern, instead quickly reequilibrating along the $r$-process path
\textemdash an interpretation that is consistent with that of
Fig.~\ref{fig:global-RW1}. To reinforce this point, we compare the
actual fission yield with the traced abundances for the
most-significantly fissioning {\nifs} nuclide, neptunium-290, in
Fig.~\ref{fig:yield-N-Np290}. While the fission yield is smoothly
distributed along $90<A<190,$ the products are eventually redistributed
throughout the second and third $r$-process peaks and long-lived
actinide isotopes, in proportion to the total abundance pattern.

Some fraction of material processed through these $\sim 9$ early-time
fission processes will eventually undergo a final late-time fission
event sometime after the free neutron abundance begins to subside.
Consequently, nuclei will be distributed according to the yields of
these final fission events without significant reprocessing by neutron
capture. Because these calculations continue to trace fission products
through all subsequent fission events, we see the formation of the
$80<A<125$ abundances in Fig.~\ref{fig:rel-all}. Our tracing
calculations suggest that, for an $r$ process proceeding via complete
fission recycling, abundance features which eventually form via
late-time fission were first processed through the fission of just a
handful of nuclear species, in this particular case the $\sim 9$ fission
processes we have identified here.

\begin{figure}[!ht]
\begin{center}
\includegraphics{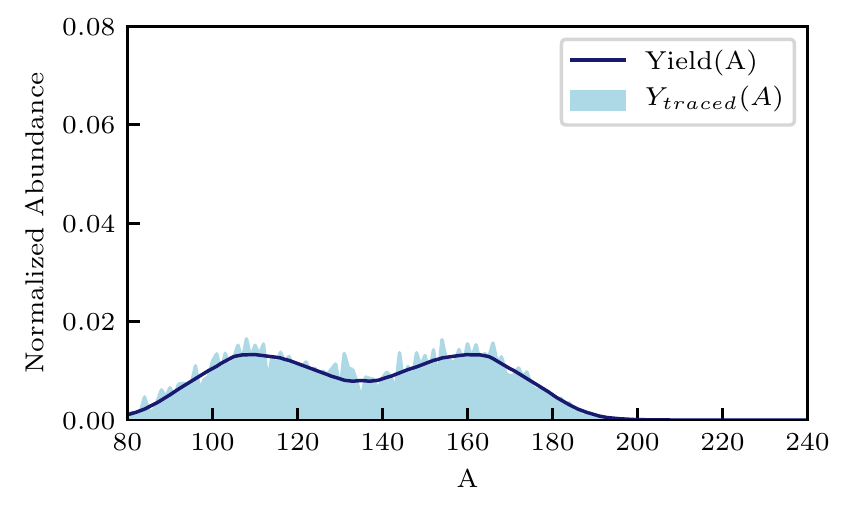}
  
\caption{\label{fig:yield-B-Bk270} Comparison of fission yield (solid
line) to final traced abundances (shaded region) for the {\bdf} of
berkelium-270, as in Fig.~\ref{fig:yield-N-Np290}. Nuclei produced by
this fission process primarily undergo a series of {\bmd s}, with
minimal effect on the distribution in mass number, $A$, compared to that
of the original yield.}
\end{center}
\end{figure}

If we consider nuclei that fission below the $10\%$ threshold used in
the preceding discussion, we find a large number of late-time fission
processes whose yields leave a static imprint on the total abundance
pattern. In Fig.~\ref{fig:yield-B-Bk270}, we compare the fission yield
to the traced abundances for one such example, the late-time {\bdfs} of
berkelium-270 (Z=97). The traced abundances, in this case, clearly
follow the fission yield, with any discrepancies arising from {\bdne}
that happens as the products decay towards stable nuclei, effectively
shifting some of the products toward lower values of $A.$

\begin{figure*}[!ht]
\begin{center}
\includegraphics[width=0.65\textwidth]{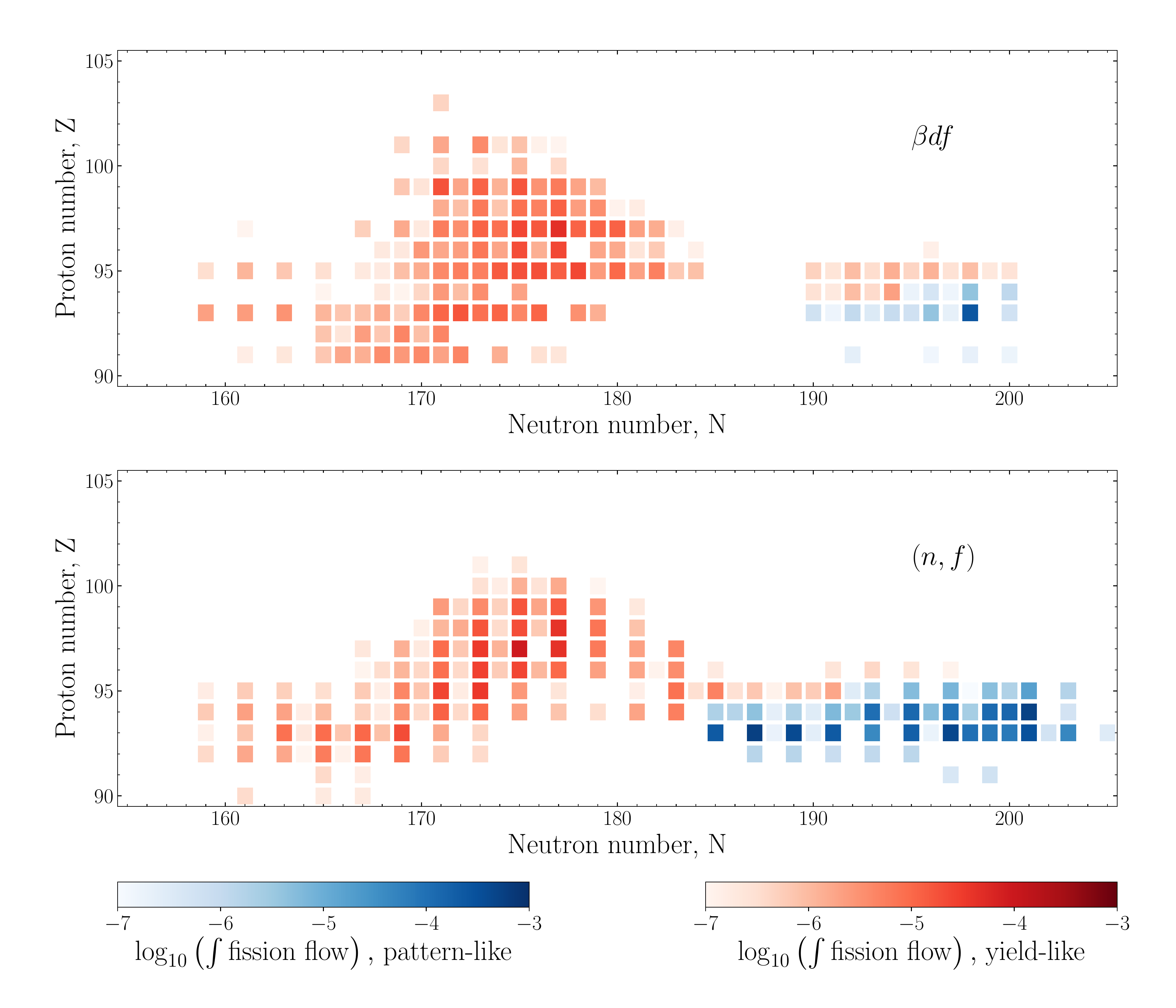}

\caption{\label{fig:fission_int_flow} Integrated $\beta$-delayed ($\beta
df$, top panel) and neutron-induced ($(n,f)$, bottom panel) fission
flows for individual nuclides during the cold dynamical ejecta
conditions of \cite{RW+Traj}, as in Fig.~\ref{fig:global-RW1}. Red color
indicates that the traced isotopic abundances are mostly similar to the
fission yield ($Y_{traced} (A) \sim \text{Yield}(A)$, as in
Fig.~\ref{fig:yield-B-Bk270}), and blue color indicates that the traced
isotopic abundances are similar to the overall pattern ($Y_{traced}(A)
\sim Y(A)$, as in Fig.~\ref{fig:yield-N-Np290}). }
\end{center}
\end{figure*}

Figure~\ref{fig:fission_int_flow} places the {\nifs} and {\bdfs} of each
nuclide into the dichotomy of \textit{pattern-like} (as in
Fig.~\ref{fig:yield-N-Np290}) and \textit{yield-like} (as in
Fig.~\ref{fig:yield-B-Bk270}) traced abundances. We begin by calculating
integrated fission flows, defined for each nuclide $i$ as
$\int \Lambda_{(n,f),i} Y_n Y_{i} \ dt$ and $\int \Lambda_{\beta df,i}
Y_{i}\ dt$ for {\nifs} and {\bdfs}, respectively; here, $\Lambda$ is as
defined in Sec.~\ref{sec:theory}. For every fission process with an
integrated fission flow in excess of $10^{-7},$ we evaluate the
functions
\begin{align}
L_{\text{pattern}} = \frac{1}{2} \sum_{A>80} | Y(A)-Y_{traced}(A) | \\
L_{\text{yield}} = \frac{1}{2} \sum_{A>80} |\text{Yield}(A) -
Y_{traced}(A)|
\end{align}
where $Y(A)$, $Y_{traced}(A)$, and $\text{Yield}(A)$ are the total
abundances, traced abundances, and fission yields, respectively, and
each is normalized such that
\begin{equation}
\label{eq:norm}
\begin{split}
& \sum_{A>80} Y(A) = 1, \\ & \sum_{A>80} \text{Yield}(A) =1,\text{ and}
\\ & \sum_{A>80}Y_{traced}(A) =1.
\end{split}
\end{equation}
In this way, $L_{\text{yield}}$ is nearly zero if the traced abundances
follow the same distribution as the corresponding fission yield.
Likewise, $L_{\text{pattern}}$ is nearly zero if the distribution of the
traced abundances follows that of the total abundances. By comparing
$L_{\text{yield}}$ to $L_{\text{pattern}},$ we may systematically
identify whether the {\nifs} and {\bdfs} of each nuclide is pattern-like
or yield-like. In Fig.~\ref{fig:fission_int_flow}, the {\nifs} and
{\bdfs} traced abundances for each nuclide are colored red if they are
yield-like and blue if pattern-like, and the shading of each indicates
integrated fission flow.

Along the $r$-process path ($Z\sim 95$ and $N\geq~185$), all of the
traced abundances are pattern-like, suggesting that these fission
products quickly reequilibrate along the existing $r$-process path. For
the remaining less neutron-rich nuclides, the traced abundances are
consistently yield-like, and their contribution to the overall isotopic
abundances are mostly in proportion to their respective fission yields.

The yield-like nuclides for {\bdfs} and {\nifs} are distributed over a
relatively large number of nuclides. Collectively, their fission
products play a significant role in shaping certain features of the
final abundance pattern, consistent with the interpretation of results
presented in Fig.~\ref{fig:global-CF20}, Fig.~\ref{fig:global-RW17}, and
Fig.~\ref{fig:global-RW1}. On the other hand, the effects of individual
fission yields are averaged out across these many different nuclides.
Indeed, in no case do any of the yield-like traced abundances constitute
more than $\sim 7\%$ of the total abundances for any value of $A.$ While
the early-time fission of the most neutron-rich nuclei tends to be more
significantly focused on only a few nuclei, these contributions tend to
be pattern-like and, therefore, similarly insensitive to fine details
the associated yields. Instead, the final nucleosynthetic outcome is
shaped by average trends in fission yields for the many nuclei populated
during the decay to stability at late stages of the $r$ process,
consistent with the results of, e.g., \cite{Goriely2015}. While these
conclusions are drawn from a specific set of nuclear data, and as shown
e.g. in \cite{Vassh+2019} different sets of data will lead to distinct
predictions for impactful fissioning nuclei, we expect these general
trends to hold.

\subsection{Tracing {\bmd s} in an $r$ process}\label{sec:betas}

Because the $r$ process involves the most neutron-rich nuclei, many of
these nuclei are difficult to study experimentally, and nucleosynthesis
simulations rely heavily on predictions from theoretical nuclear models.
Beyond the limits of experimental data, theoretical predictions for
these nuclei diverge \cite{Erler2012,McDonnell2015,Neufcourt2020},
introducing a significant source of uncertainty in $r$ process
nucleosynthesis simulations
\cite{Mumpower+2016,Surman+2016,Martin+2016,Sprouse+DFT}. Experimental campaigns at current and upcoming facilities such as CARIBU
\cite{Scielzo2012,Yee2013,Scielzo2014,Wang2020,Hirsh+2016,Orford+2018}
and the $N=126$ Factory \cite{Savard2020} at ATLAS, IGISOL at
Jyv{\"a}skyl{\"a} \cite{Kankainen+2013,Vilen+2018}, ISOLDE at CERN
\cite{Lunney+2017}, TITAN at TRIUMF \cite{Lascar+2017}, RIKEN
\cite{Moon2017,Fukuda2018,Delgado2019}, GSI/FAIR
\cite{Stocker2015,Folch2017,Gottardo2019}, and FRIB \cite{FRIB} are
approaching nuclei of interest to the $r$ process.

In this context, it will be especially important to identify which
nuclei are of critical importance to understanding and constraining
$r$-process nucleosynthesis simulations. Here, we demonstrate how our
nucleosynthesis tracing technique can help to accomplish this task. We
focus specifically on only one category of nuclear data on which $r$
process nucleosynthesis simulations critically depend, namely {\bmd}
properties for neutron-rich nuclei. Nuclear {\bmd} is the transmutation
process responsible for moving neutron rich nuclei towards heavier
elements during the $r$ process. In addition to controlling the number
retained at `waiting points' associated with closed neutron shells,
{\bmd} can also compete with $(n,\gamma)$ and $(\gamma,n)$ reactions to
adjust the nuclear abundances produced during the $r$ process for as
long as these reaction channels remain active (see, e.g.,
\cite{Surman1997,Surman2001,Wanajo+2007,
Arcones2011,Mumpower2012,Mumpower2014}).
We focus on identifying which of these {\bmd s} an $r$ process most
strongly depends.

To simplify the interpretation of our results, we select a parameterized
neutron star merger wind in which fission does not participate as an
active process during nucleosynthesis, with parameters $s/k_B=50,$ $\tau
= 50$~ms, and $Y_e=0.25$. For each nuclide with $40 \leq Z \leq 80$
populated at any point during nucleosynthesis, we perform a tracing
calculation for its {\bmd}. The resulting calculation indicates the
relative fraction of each abundance with a history involving the {\bmd}
under consideration.

\begin{figure}[!ht]
\begin{center}
\includegraphics{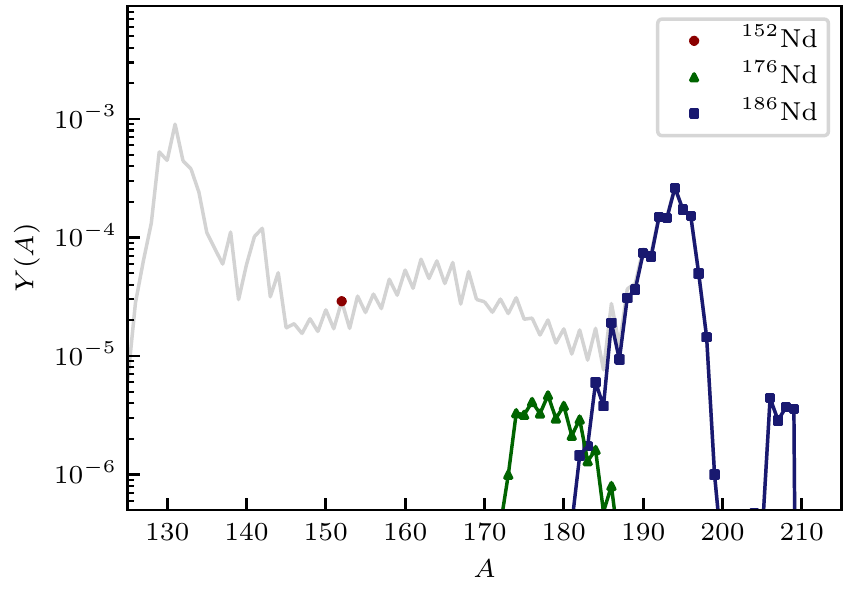}
  
\caption{\label{fig:beta-rel-select} Traced isotopic abundances for the
{\bmd} of a selection of neodymium isotopes (neodymium-152, red circle;
neodymium-176, green triangles; and neodymium-186, blue squares). The
total abundances are given by the gray line for comparison.
Neodymium-152 is populated as nuclei are decaying to stable nuclei, and
the entirety of abundances for $A=152$ have undergone this {\bmd}.
Neodymium-186 lies on the $r$-process path, and all nuclei with $A \geq
186$ have undergone this {\bmd}. Neodymium-176 represents an
intermediate case, where some fraction, $\sim 10\%$, of nuclei with $A
\sim 176$ having been produced by this {\bmd}. }
\end{center}
\end{figure}

The traced {\bmd s} can be roughly sorted into three distinct, yet
physically intuitive, categories. For nuclides nearest stability, their
{\bmd s} occur well after the free neutron abundance has been exhausted,
and so these preserve the mass number $A$ with respect to the final
pattern. Along the $r$-process path, the most neutron-rich isotopes
populated during an $r$ process, the bulk of of all heavier nuclei will
proceed through these nuclei via {\bmd}. As a result, the traced
abundances will reproduce nearly the entire pattern for all larger
values of $A.$ Finally, one can imagine an intermediate case, where
nuclei begin to fall back towards stability as a result of decreasing
free neutron abundances but may still participate in some degree of
neutron capture. In Fig.~\ref{fig:beta-rel-select}, we show an example
of each regime for a selection of neodymium ($Z=60$) isotopes. In the
case of neodymium-152, all of the abundances in the final overall
pattern have participated in this {\bmd} while decaying back to
stability after the completion of the $r$ process. For neodymium-186,
which lies along the $r$ process path, the majority of populated nuclear
species with $A\geq 186$ have participated in this particular {\bmd}.
Finally, we highlight the intermediate case with neodymium-176, lying
between stability and the $r$-process path, where some fraction of
abundances for multiple nearby $A$ have a history involving this {\bmd}.

\begin{figure*}[!ht]
\begin{center}
\includegraphics[width=0.65\textwidth]{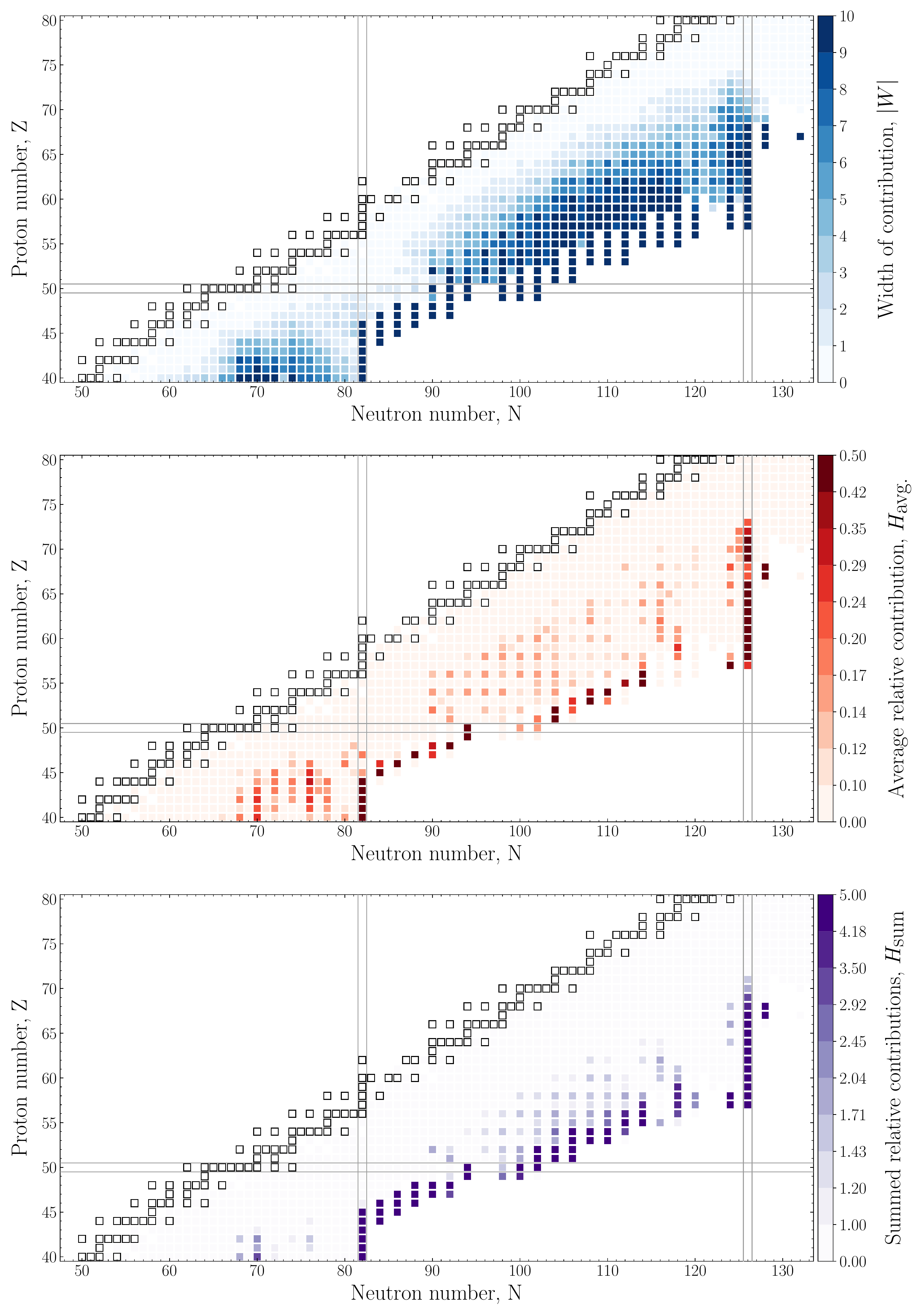}

\caption{\label{fig:betas-NZ} Average trends in the traced abundances of
individual {\bmd}s for elements $40<Z<80.$ The top panel shows the
`width' ($|W|$) of the traced abundances, as defined in the text. The
middle panel indicates the average relative contribution to total
abundances contained within the width ($H_\text{avg.}$) . The bottom
panel gives the the sum of relative contributions to the width
($H_\text{sum}$), which may be interpreted as the product of the top and
middle panels.}
\end{center}
\end{figure*}

In Fig.~\ref{fig:betas-NZ}, we quantify average trends that arise in
these tracing network calculations. For each traced pattern, we define
the set
\begin{equation}\label{eq:W}
W = \left\{A' > A \mid Y_{\text{traced}}(A')/Y(A') > 1\%    \right\}  
\end{equation}
where $A$ is the mass number of the traced {\bmd} parent nucleus. The
set $W$ represents the values of $A$ for which the traced abundance
represent at least $1\%$ of the final abundance. We refer to the
cardinality of the set $W$ as the \textit{width} of the traced pattern.
By only considering $A'>A,$ we omit any contributions to the final
pattern that occur during decay back to stability following the $r$
process. As a result, the width is restricted to contributions that are
dynamically involved in the $r$ process.

In the top panel of Fig.~\ref{fig:betas-NZ}, we report the width of each
of our tracing calculations. For each element, the width is greatest
along the $r$-process path, since the vast majority of abundances of
heavier nuclides are produced along this path. Near stability, the width
collapses to $0$ because all subsequent nucleosynthesis strictly follows
a series of {\bmd}s that preserve mass number $A$, which is omitted from
the set $W$ as we have constructed it. In the intermediate region, there
is a smooth transition from larger to smaller widths, with values
ranging from 2-10 for a relatively large number of nuclides lying away
from the $r$ process path.

It is also instructive to consider the average contribution of a
particular {\bmd} to the total abundance pattern. We define an
additional metric that attempts to provide this insight, defined as
\begin{equation}\label{eq:Favg}
H_{\text{avg.}} = \frac{1}{|W|}\cdot\sum\limits_{A'\in W}
Y_{traced}(A')/Y(A')
\end{equation} 
where $W$ is the same as in Eq.~\ref{eq:W} and $|W|$ is the width. The
value of $H_{\text{avg.}}$ can be understood as the relative
contribution of a particular {\bmd}, on average, to nuclides contained
within the width of the contribution. A large value represents
significant contributions to the entirety of the width of the traced
pattern, while smaller values correspond to less significant
contributions.

Analogous to this metric is the summed relative contributions, given by\begin{equation}\label{eq:Fsum}
H_{\text{sum}} = \sum\limits_{A'\in W} Y_{traced}(A')/Y(A') 
\end{equation}
Large values in this metric indicate a relatively large width together
with significant contributions to overall abundances to the same width.

The values of $H_{\text{avg.}}$ and $H_{\text{sum}}$ are shown in the
middle and bottom panels of Fig.~\ref{fig:betas-NZ}, respectively. As
with the widths shown in the top panel, the largest values in each
metric lie along the path for the same reasons previously discussed.
However, in the intermediate region lying between the $r$-process path
and decay-to-stability nuclides, we can further constrain the list of
`impactful' {\bmd}s for these conditions. Many of the {\bmd}s with
relatively wide contributions to the abundance pattern have
comparatively weak contributions, less than $10\%,$ and may reasonably
be considered less important in determining the final abundances
overall.

As can be observed in the bottom panel, {\bmd} for nuclides nearer the
$r$-process path have sufficiently wide contributions affecting a larger
region of the abundance pattern. For nuclides in the intermediate region
but nearer stability, there can be still-significant contributions, but
these contributions are focused on more constrained regions of the
abundance pattern, as they have a significant value for
$H_{\text{avg.}}$ but smaller values for the width and, consequently,
$H_{\text{sum}}.$
      
Clearly, {\bmd} rates along the $r$-process path dominate all three of
the metrics considered in this section. However, by applying the
nucleosynthesis tracing framework, it becomes clear that a large number
of {\bmd} rates for less neutron-rich nuclides can also significantly
influence nucleosynthesis in $r$-process environments.

As a final caution, we point out that these results strongly depend on
the astrophysical conditions and the theoretical nuclear models used to
supplement available experimental data. Changes to either are liable to
affect the $r$-process path and the onset of fission recycling, in
addition to other possible complications. We propose here only the
method by which more robust analyses may proceed in future works.
However, we do anticipate the general result to hold; namely, {\bmd}
rates of many nuclides less neutron-rich than the $r$-process path are
still important in determining nucleosynthesis. We emphasize the
importance of future experiments that measure the {\bmd} rates (or other
properties, such as nuclear masses) for these nuclides, even if the most
neutron-rich nuclei remain out of reach for the foreseeable future.

\section{Conclusion}
The most complex examples of nucleosynthesis involve thousands of
nuclear species connected by many tens of thousands of nuclear
transmutation processes. Owing partly to this complexity, as well as to
the generally dynamic and nonlinear nature of nucleosynthesis, it is
often difficult to study subsets of nuclear properties in isolation from
a nucleosynthetic system as a whole. In this work, we develop our
nucleosynthesis tracing framework, which may be applied to partly
address this problem.

Beginning with the system of coupled differential equations constituting
a standard nuclear reaction network, we frame our tracing framework as
the separation of nuclear abundances into two populations, those of
\textit{traced} and \textit{untraced} abundances. Furthermore, we allow
each transmutation process in a network calculation to be categorized by
the way it maps reactants and products between the traced and untraced
populations. Within this schema, we derive an additional set of
differential equations that model the evolution of the traced
abundances. These additional equations are coupled to those of the
standard network; when solved together, one obtains a quantitative
description of how products from specific nuclear transmutation
processes participate in all subsequent nucleosynthesis.

We implement our tracing framework into a new version of our PRISM
reaction network, PRISM\textsuperscript{tr}, and comment on several
details regarding this implementation. Notably, the tracing network
equations are structurally similar to those of a standard reaction
network; therefore, numerical techniques commonly used to solve the
standard set of network equations are expected to be equally well-suited
for solving the tracing network equations.

In order to demonstrate some of the nucleosynthesis analyses enabled by
our tracing framework, we perform tracing network calculations using
PRISM\textsuperscript{tr} to study fission and {\bmd} as they occur in
the $r$ process of nucleosynthesis.

Our application of tracing to distinct fission channels can quantify the
influence of each channel on forming the final abundance pattern, with
qualitative results consistent with investigations of fission in the $r$
process found in the literature. These same calculations offer insight
into the extent of fission recycling in an $r$-process simulation
\textemdash in particular, they allow for the quantitative distinction
between \textit{complete} and \textit{incomplete} fission recycling in
the $r$ process.

When the tracing framework is applied to individual fission
reactions/decays, we find the fission of a relatively small number
of nuclear species along the $r$-process path drives fission recycling.
The fission yields of these nuclei have limited impact on the final
abundance pattern since the fission products undergo subsequent neutron
captures and are redistributed throughout the network. The shape of the
final abundance pattern is instead determined by the product yields of
the many nuclear species that fission as the $r$-process path moves back
to stability upon neutron exhaustion. Thus, average trends in fission
yields for a large number of nuclei are needed to characterize
$r$-process nucleosynthesis.

Additionally, we apply nucleosynthesis tracing to perform a
comprehensive examination of the {\bmd} of nuclei with atomic number
$40\leq Z\leq 80.$ We quantify the relative contribution of each {\bmd}
to $r$-process nucleosynthesis in neutron star merger wind-like
conditions, and we define several metrics that may be useful for
characterizing the nature of these contributions.

Finally, we strongly emphasize that our fission and {\bmd} results
depend on the astrophysical conditions and underlying nuclear models
used for this study; a thorough investigation of these dependencies,
together with a more comprehensive examination of the numerous and
varied nuclear properties entering into $r$-process nucleosynthesis
calculations, is intended for future work.

While we limit the present work to $r$-process applications, we note
that our tracing framework \textemdash as we have presented it
\textemdash is in no way limited to $r$-process nucleosynthesis, and it
may be readily applied to any process for which nuclear reaction
networks are appropriate. Indeed, the defining principles of our tracing
framework can be naturally adapted to applications outside of
nucleosynthesis entirely, e.g., chemical reaction networks.

\section{Acknowledgements}
This work was supported in part by the US Department of Energy under
Contract No.\ DE-FG02-95-ER40934, the topical collaboration Fission In
R-process Elements (FIRE) Contract No.\ DE-AC52-07NA27344, and SciDAC
Contract No.\ DE-SC0018232.
MM was supported by the US Department of Energy through Los Alamos
National Laboratory and by the Laboratory Directed Research and
Development program of Los Alamos National Laboratory under project
number 20190021DR.
Los Alamos National Laboratory is operated by Triad National Security,
LLC, for the National Nuclear Security Administration of U.S.\
Department of Energy (Contract No.\ 89233218CNA000001). T.S. was
supported in part by the Los Alamos National Laboratory Center for Space
and Earth Science, which is funded by its Laboratory Directed Research
and Development program under project number 20180475DR.

\appendix*
\label{appendix}
\section{Some comments on our reaction network formalism}
The notation we adopt in our construction of the nuclear reaction
network equations was chosen to simplify the expressions used in the
derivation and statement of the tracing network equations. However, this
notation differs from more commonly adopted forms, such as that used in
\cite{Hix+1999}. Here, we relate our notation to this more common
version.

Beginning with Eq.~12 from \cite{Hix+1999}, the time derivative
$\dot{Y_i}$ of each nuclear abundance $Y_i$ is given by
\begin{equation}\label{eq:hix_rxn}
\begin{split}
\dot{Y_i} = & \sum_{j} \mathcal{N}^i_j \lambda_j Y_j + \sum_{j,k}
\mathcal{N}_{j,k}^i \rho N_A \langle j,k \rangle Y_j Y_k \\
+& \sum_{j,k,l} \mathcal{N}_{j,k,l}^{i} \rho^2 {N_A}^2 \langle j,k,l
\rangle Y_j Y_k Y_l ,
\end{split}
\end{equation}
where each sum is taken over the one-, two-, and three-body reactions in
which species $i$ is either created or destroyed. Here, $\lambda_j$ is a
decay rate for species $j,$ $\rho$ is the density, $N_A$ is Avogadro's
number, $\langle j,k \rangle$ is the thermal reaction cross section for
a reaction between species $j$ and $k$, and $\langle j,k,l \rangle$ is
the thermal cross section for a reaction between species $j$, $k$, and
$l.$ The $\mathcal{N}_j^i,$ $\mathcal{N}_{j,k}^i,$ and
$\mathcal{N}_{j,k,l}^i$ are numerical factors that correctly count the
number of species consumed or produced in each reaction, defined as
\begin{equation}\label{eq:hix_n}
\begin{split}
\mathcal{N}_j^i =& N_i, \\ 
\mathcal{N}_{j,k}^i =& N_i / \prod_{m=1}^{n_m} | N_{j_m} | !,\text{ and}
\\
\mathcal{N}_{j,k,l}^i =&  N_i / \prod_{m=1}^{n_m} | N_{j_m} | !, \\
\end{split}
\end{equation}
with $N_i$ giving the number of species $i$ produced (positive) or
consumed (negative) by a reaction or decay, and the product in the
denominator run over all species consumed by a reaction and corrects for
overcounting a reaction involving identical reactants.

In relation to the terminology and notation we use in
Sec.~\ref{sec:theory}, each term in Eq.~\ref{eq:hix_rxn} corresponds to
a unique process, $p,$ in a network. The processes are grouped into
one-, two-, and three-body processes in the first, second, and third
sums, respectively. In the case of one-body process, then, we
have the associations
\begin{itemize}
\item $\Lambda_p = \lambda_j,$
\item $\alpha_p(i) = 1$ if $i$ is a reactant of $p$ and 0 otherwise, and\item $\beta_p(i) = |N_i|$ if $i$ is a product of $p$ and 0 otherwise.
\end{itemize}
For two-body processes, the associations are given by
\begin{itemize}
\item $\Lambda_p = \rho N_A \langle j,k \rangle \cdot \frac{
\mathcal{N}^i_{j,k}}{N_i},$
\item $\alpha_p(i) = |N_i|$ if $i$ is a reactant of $p$ and 0 otherwise,
and
\item $\beta_p(i) = |N_i|$ if $i$ is a product of $p$ and 0 otherwise.
\end{itemize}
Finally, for three-body processes, we have
\begin{itemize}
\item $\Lambda_p = \rho^2 {N_A}^2 \langle j,k,l \rangle \cdot \frac{
\mathcal{N}^i_{j,k,l}}{N_i},$
\item $\alpha_p(i) = |N_i|$ if $i$ is a reactant of $p$ and 0 otherwise,
and
\item $\beta_p(i) = |N_i|$ if $i$ is a product of $p$ and 0 otherwise.
\end{itemize}
In all cases, the sets $\mathcal{R}_p$ and $\mathcal{P}_p$ simply
collect the reactants and products of the process $p$, which we use to
make explicit the limits of the summations and products used in
Eq.~\ref{eq:traditional_net}.

By writing the one-, two-, and three-body terms separately, collecting
the positive and negative terms of each into a common summation, and
performing the substitutions defined above, Eq.~\ref{eq:traditional_net}
can be rewritten as Eq.~\ref{eq:hix_rxn}.

\newpage
\bibliographystyle{unsrt}
\bibliography{refs}

\end{document}